\documentclass[11pt,a4paper]{article}
\usepackage{amstex}
\newcommand{\Nti}{\ensuremath{\underset{\!\!\sim}{N}}}

\begin{document}
\title{\textbf{Metric Versus Ashtekar Variables in Two Killing Field Reduced 
Gravity}}
\makeatletter
\author{Marco Zagermann\thanks{Present address: Physics Department, Penn 
State University 104 Davey Lab., University Park, PA 16802, USA. 
E-mail: zagerman@phys.psu.edu}\\ \small{II. Institut f{\"u}r Theoretische 
Physik, Universit{\"a}t Hamburg} \\ \small{Luruper Chaussee 149, D-22761 
Hamburg, Germany}\\}
\makeatother 
\maketitle
\begin{abstract}
The relation between the $SL(2,\mathbb{R})/SO(2)$- and the 
$SL(2,\mathbb{C})$-chiral model that naturally arise within the metric 
respectively the Ashtekar formulation of two Killing field reduced Einstein 
gravity is revealed.  Both chiral models turn out to  be completely equivalent
even though the transition from the coset- to the $SL(2,\mathbb{C})$-model is 
accompanied by a disappearance of the non-ultralocal terms in the Poisson 
brackets.
\end{abstract}

Among the various toy models which have been investigated in order to discuss 
in a simplified context some characteristic features of the canonical approach
to quantum gravity, the two Killing field reduction of general relativity 
plays a particularly important r\^{o}le. Indeed, it belongs to the simplest 
reductions of full $4D$ general relativity that still describe nonlinear 
genuine field theories without a background metric. In this letter we will 
compare the outcomes of two different formulations of this reduction:

In the conventional metric picture, the symmetry reduction naturally leads to 
a $2D$ dilaton-gravity coupled $SL(2,\mathbb{R})/SO(2)$-$\sigma$-model. The 
underlying Lagrangian induces Poisson brackets with  non-ultralocal terms 
(i.e. terms containing derivatives of $\delta$-functions) between the 
corresponding $\sigma$-model currents.

On the other hand, performing the symmetry reduction  within the Ashtekar 
formulation \cite{Ash86Ash87}, naturally results in a description in terms of 
a generalized $SL(2,\mathbb{C})$-chiral model with completely ultralocal 
Poisson brackets \cite{Hus96,HS89}. As for the  equations of motion, however, 
one encounters a striking similarity to the 
$SL(2,\mathbb{R})/SO(2)$-formulation.

To understand the surprising coexistence of these two chiral models we will 
translate them into each other, verify  whether they are equivalent and 
observe the mechanism that causes the differences between them (while 
preserving the similarities). The result will be interesting in its own right,
but should also shed some light on the relation of the non-local charges 
constructed in \cite{KorSam96b} and  \cite{Hus96}.

To begin with, let us briefly recapitulate the symmetry reduction using the 
ordinary metric formalism:

A spacetime possessing two commuting spacelike and two-surface orthogonal  
Killing vector fields admits the choice of a local coordinate system 
$(t,x,y,z)$ such that the Killing vectors are given by 
$\frac{\partial}{\partial y}$ and $\frac{\partial}{\partial z} $ and the  
metric  $G_{MN}$  attains the following form
\begin{equation}\label{sum0}
G_{MN} = G_{MN}(t,x)  =
\left(\begin{array}{cc}
\rho^{-1/2}e^{2k}\eta_{\mu\nu} & 0\\
0 &\rho  g_{\bar{m}\bar{n}}
\end{array}\right),
\end{equation}
where $M,N,\ldots \in  \{t,x,y,z\};\quad$  $\mu,\nu,\ldots \in  \{t,x\};\quad$
$\bar{m},\bar{n},\ldots \in  \{y,z\}\quad$;
$\det g_{\bar{m}\bar{n}}   = 1$; 
$\eta_{\mu\nu} = \textrm{diag}\,(-1,+1)$ and $k$ and $\rho$ are some real 
functions.
Note that $g=g^{t}$, which together with $\det g = 1$ means that $g$ can be 
viewed as $SL(2,\mathbb{R})/SO(2)$-valued.
The Einstein equations imply the wave equation 
$\eta^{\mu\nu}\partial_{\mu}\partial_{\nu}\rho =0$
so that in the case of timelike $\partial_{\mu}\rho$ (i.e. for example in 
Gowdy's $T^{3}$-model \cite{Gow74}) the additional coordinate fixing $t=\rho$ 
(the non-stationary analog of Weyl's canonical coordinates)
can be used to obtain a simplified version of the remaining Einstein equations:
\begin{eqnarray}
\partial_{t}k =   \frac{1}{8t} \textrm{tr}({J_{0}}^{2}+{J_{1}}^{2})&,&\quad
\partial_{x}k  = \frac{1}{4t} \textrm{tr}(J_{0}\cdot J_{1})\nonumber\\
\partial_{t}J_{0}-\partial_{x}J_{1} & = &0,\label{sum1}
\end{eqnarray}
where the currents $J_{0},\  J_{1} $ are defined by
\begin{displaymath}
J_{0}:=t g^{-1}\partial_{t}g,\quad J_{1}:=t g^{-1}\partial_{x}g
\end{displaymath}
and therefore obey the following integrability condition
\begin{equation}\label{sum2}
\partial_{t}J_{1}-\partial_{x}J_{0}-\frac{1}{t}[J_{1},J_{0}]-\frac{1}{t}J_{1}=
0.
\end{equation}
Eqs. (\ref{sum1}) and (\ref{sum2}) can be written as the compatibility 
conditions of a linear system \cite{MBZ78} with a spacetime (ie. $(t,x)$-) 
dependent  spectral parameter. It is essentially the $\frac{1}{t}J_{1}$-term 
in (\ref{sum2}) which requires this spacetime dependence.

To obtain the Poisson brackets between the currents $J_{\mu}$ from the 
(relevant part of the) symmetry reduced Lagrangian
\begin{equation}\label{sum2a}
\mathcal{L}=Ct^{-1} \: \textrm{tr}[J_{0}^{2} - J_{1}^{2}],
\end{equation}
where $C$ is a constant,   one has to take into account  the symmetry 
$(g=g^{t})$ and the unimodularity of $g$. One (and probably the most 
economical) way to do this, is to use the formalism described in 
\cite{KorSam96b}. An equivalent (less abstract but more tedious) method is 
to  parametrize $g$ by two independent fields $f$ and $\Phi$:
\begin{equation}\label{sum2b}
g=\left(\begin{array}{cc}
f^{-1} & \Phi f^{-1}\\
\Phi f^{-1} & f+ \Phi^{2}f^{-1}\end{array}\right).
\end{equation}
so that $\mathcal{L}$ becomes 
\begin{equation}\label{sum2c}
\mathcal{L}=2Ctf^{-2}[(\partial_{t}f)^{2} + (\partial_{t}\Phi)^{2}-
(\partial_{x}f)^{2}-(\partial_{x}\Phi)^{2}].
\end{equation}
Either way, the result is: 
\begin{eqnarray}
\{J_{0}(x)\stackrel{\otimes}{,}J_{0}(x')\} & = & 
\frac{1}{4C}[\Pi,\mathbf{1}\otimes J_{0}]\delta(x-x')\label{sum3a}\\
\{J_{1}(x)\stackrel{\otimes}{,}J_{0}(x')\} & = & 
\frac{1}{4C}[\Pi,\mathbf{1} \otimes J_{1}]\delta(x-x')\nonumber\\
& &\!\!\!\!+\frac{t}{4C} (2\Pi-\mathbf{1}\otimes\mathbf{1}  +  
\varepsilon g \otimes g^{-1} \varepsilon)(x)\cdot 
\partial_{x} \delta(x-x')\label{sum3b}\\
\{J_{1}(x)\stackrel{\otimes}{,}J_{1}(x')\} & = & 0,\label{sum3c}
\end{eqnarray}
where $\{A\stackrel{\otimes}{,} B \}_{\alpha \beta, \gamma\delta}\equiv 
\{A_{\alpha \beta},  B_{\gamma\delta}\}$ and $\Pi$ denotes the permutation 
operator on $\mathbb{C}^{2}\otimes\mathbb{C}^{2}$
\begin{displaymath}
\Pi =\left(\begin{array}{cccc}
1 & 0 & 0 & 0\\
0 & 0 & 1 & 0\\
0 & 1 & 0 & 0\\
0 & 0 & 0 & 1
\end{array}\right)\Leftrightarrow \Pi_{\alpha \beta, \gamma\delta}= 
\delta_{\alpha\delta}\delta_{\beta\gamma},
\end{displaymath}
whereas the $(2\times 2)$-matrix $\varepsilon$ is given by
\begin{equation}\label{sum3d}
\varepsilon = \left(\begin{array}{cc}
0 & 1\\
-1 & 0
\end{array}\right).
\end{equation}
Normally, non-ultralocal terms like those in (\ref{sum3b}) destroy the 
canonical formalism and lead to unresolvable ambiguities in the corresponding 
quantum theory. In the case at hand, however, these terms  combine, roughly 
speaking, with the spacetime dependence of the spectral parameter of the 
linear system to yield unambigious results at the level of transition matrices
\cite{KorSam96b}. This observation has been used to identify the conserved 
non-local charges that generate the Geroch
group \cite{Ger72} with respect to the Poisson structure 
(\ref{sum3a})-(\ref{sum3c}) \cite{KorSam96b} and to perform a consistent 
quantization in the case of cylindrical symmetry \cite{KorSam97a}.

We now  sketch how the above-mentioned generalized $SL(2,\mathbb{C})$-chiral 
model emerges within the Ashtekar formulation, thereby fixing our notation:

Indices from the middle (beginning) of the alphabet  will denote  spacetime 
(internal) indices:
\begin{eqnarray*}
M,N,\ldots \in  \{t,x,y,z\};\quad  m,n,\ldots &\in & \{x,y,z\};\quad
\bar{m},\bar{n},\ldots \in  \{y,z\}\\
A,B,\ldots \in  \{0,1,2,3\};\quad
a,b,\ldots &\in & \{1,2,3\};\quad
\bar{a},\bar{b},\ldots \in  \{2,3\}.
\end{eqnarray*}
The inverse densitized dreibein $\tilde{e}_{a}^{m}=e_{a}^{m}\det(e_{n}^{b})$ 
and the components of the Ashtekar connection $A_{ma}$  satisfy the 
fundamental Poisson brackets
\begin{displaymath} 
\{\tilde{e}_{a}^{m}(x),A_{nb}(x')\}  = 
-i\, \delta_{ab}\delta_{n}^{m}\delta^{(3)}(x-x')
\end{displaymath}
and are subject to the first class constraints (with $a,b,\ldots$ raised with 
$\delta^{ab}$)
\begin{eqnarray*}
\mathcal{H}\,\, & :=& \varepsilon^{abc} 
F^{a}_{mn}\tilde{e}^{mb}\tilde{e}^{nc}\approx 0 \label{As3b}\\
\mathcal{C}_{m} & :=& F^{a}_{mn}\tilde{e}^{na}\approx 0\label{As3c}\\
\mathcal{G}^{a} & :=& D_{m}\tilde{e}^{ma}\approx 0\label{As3d}
\end{eqnarray*}
leading to the total Hamiltonian (without surface terms)
\begin{displaymath}
H_{\textrm{tot}}\, (\Nti,N^{m},\Lambda_{a})=\int\nolimits_{\Sigma} d^{3}x 
\left(\frac{1}{2}\Nti\mathcal{H}+iN^{m}\mathcal{C}_{m}+
\Lambda_{a}\mathcal{G}^{a}\right).
\end{displaymath}
As usual, $\Nti=N/\det(e_{m}^{a})$, $N^{m}$,  $D_{m}$ and $F_{mna}$ are the 
densitized lapse function, the shift vector, the covariant derivative with 
respect to $A_{ma}$ and the corresponding field strength, respectively.

The reduction can now be divided into four main steps (see 
\cite{Hus96,HS89,Men97} for details):\\
(i)\emph{A first gauge fixing:} We use adapted coordinates such that the two 
Killing vectors are given by the coordinate vector fields 
$\frac{\partial}{\partial y}$ and $\frac{\partial}{\partial z} $,  implying 
the $(y,z)$-independence of all phase space variables. Subsequent imposition 
of  the partial gauge fixing conditions $\tilde{e}_{\bar{a}}^{x}=
\tilde{e}_{1}^{\bar{m}}=0$ breaks part of the $SO(3)$- and diffeomorphism 
invariance, while solving the resulting second class constraints requires 
$A_{\bar{m}1} =A_{x\bar{a}} = 0$  \cite{HS89}  and leaves us with a reduced 
phase space consisting of the canonical pairs 
\begin{eqnarray*}
A:=A_{x1} & , & E:=\tilde{e}_{1}^{x}\\
A_{\bar{m}\bar{a}} & , & \tilde{e}_{\bar{a}}^{\bar{m}}
\end{eqnarray*}
and the remaining first class constraints
\begin{eqnarray*}
G & := & \partial_{x}E + \varepsilon^{\bar{a}\bar{b}}A_{\bar{m}\bar{a}}
\tilde{e}_{\bar{b}}^{\bar{m}} \approx 0 \label{As6b}\\
C & := & A\partial_{x}E - \tilde{e}_{\bar{a}}^{\bar{m}}
\partial_{x}A_{\bar{m}\bar{a}}\approx 0 \label{As6c}\\
H & := & -2\varepsilon^{\bar{a}\bar{b}}F_{x\bar{m}\bar{a}}
\tilde{e}_{\bar{b}}^{\bar{m}}E + F_{\bar{m}\bar{n}3}
\tilde{e}_{\bar{a}}^{\bar{m}}\tilde{e}_{\bar{b}}^{\bar{n}}
\varepsilon^{\bar{a}\bar{b}}\approx 0,\label{As6d}
\end{eqnarray*}
where $\varepsilon^{\bar{a}\bar{b}}=-\varepsilon^{\bar{b}\bar{a}},\quad 
\varepsilon^{23}=+1$. \\
(ii)\emph{New variables:} Defining the $(2\times 2)$-matrices
\begin{eqnarray*}
B_{0} & = & (B_{0})_{\bar{n}\bar{m}}:= i\, K_{\bar{m}}^{\bar{n}}:=i\, 
A_{\bar{m}\bar{a}}\tilde{e}_{\bar{a}}^{\bar{n}}\label{As10a}\\
B_{1} & = & (B_{1})_{\bar{n}\bar{m}}:= J_{\bar{m}}^{\bar{n}}:= 
\varepsilon^{\bar{a}\bar{b}} A_{\bar{m}\bar{a}}
\tilde{e}_{\bar{b}}^{\bar{n}},\label{As10b}
\end{eqnarray*}
one finds the following Poisson brackets
\begin{eqnarray*}
\{B_{0}(x)\stackrel{\otimes}{,}B_{0}(x')\} & = & 
[\Pi,B_{0}\otimes \mathbf{1}]\delta(x-x')\label{As11a}\\
\{B_{1}(x)\stackrel{\otimes}{,}B_{0}(x')\} & = & 
[\Pi,B_{1}\otimes \mathbf{1}]\delta(x-x')\label{As11b}\\
\{B_{1}(x)\stackrel{\otimes}{,}B_{1}(x')\} & = & 
[\Pi,B_{0}\otimes \mathbf{1}]\delta(x-x')\label{As11c}.
\end{eqnarray*}
This  together with $\{B_{\mu},K\}=\{B_{\mu},J\} = 0$, where 
$K:= K_{\bar{m}}^{\bar{m}}$ and $J:= J_{\bar{m}}^{\bar{m}}$, implies that 
the corresponding traceless (i.e. $sl(2,\mathbb{C})$-valued) parts 
\begin{eqnarray*}
A_{0} & := & B_{0}- \frac{1}{2} \textrm{tr}B_{0}\cdot \mathbf{1}= B_{0}- 
\frac{1}{2} i\, K\cdot \mathbf{1}\label{As16a}\\
A_{1} & := & B_{1}- \frac{1}{2} \textrm{tr}B_{1}\cdot \mathbf{1}= B_{1}- 
\frac{1}{2}  J\cdot \mathbf{1}.\label{As16b}
\end{eqnarray*}
satisfy the Poisson brackets 
\begin{eqnarray}
\{A_{0}(x)\stackrel{\otimes}{,}A_{0}(x')\} & = & [\Pi,A_{0}\otimes 
\mathbf{1}]\delta(x-x')\label{sum5a}\\
\{A_{1}(x)\stackrel{\otimes}{,}A_{0}(x')\} & = & [\Pi,A_{1}\otimes 
\mathbf{1}]\delta(x-x')\label{sum5b}\\
\{A_{1}(x)\stackrel{\otimes}{,}A_{1}(x')\} & = & [\Pi,A_{0}\otimes 
\mathbf{1}]\delta(x-x').\label{sum5c}
\end{eqnarray}
(iii)\emph{The (nontrivial) equations of motion:} The time-dependence of 
$A_{\mu}$ follows from $\partial_{t}A_{\mu}=\{A_{\mu},H_{tot}(\Nti)\}$ 
with\footnote{The other two constraints $C$ and $G$, which normally would also
appear in  $H_{\textrm{tot}}$ with Lagrange multipliers $iN^{x}$ and 
$\Lambda$, need not  be considered here, because $A_{\mu}$ commutes with $G$ 
and an additional gauge fixing (see (iv)) will require  $N^{x} = 0$.}  
$H_{\textrm{tot}}\, (\Nti)=\int dx \left(\frac{1}{2}\Nti H\right)$, yielding
\begin{eqnarray}
\partial_{t}A_{0} & = & \partial_{x}(\Nti E A_{1})\label{sum9a}\\
\partial_{t}A_{1} & = & \partial_{x}(\Nti E A_{0})-\Nti [A_{1},A_{0}].
\label{sum9b}
\end{eqnarray}
(iv)\emph{Further gauge fixing:} In order to fix the coordinates $t$ and $x$ 
completely, we demand
$E=t$,
which requires 
$\Nti = \frac{i}{EK} = \frac{i}{tK}$
for consistency, while the condition
$K= i(=\textrm{const.})$
implies $\partial_{x}N^{x}=0$ and  fixes the $x$-coordinate up to 
transformations  $x\rightarrow x+f(t)$ \cite{Hus96}. This freedom can then be 
used to absorb
$N^{x}$ by making the choice $f(t)=\int^{t}_{t_{0}} dt' N^{x}(t')$ 
\cite{Men97}.
Since $A_{\mu}$ commutes with $E$ and $K$, these gauge fixings do not alter 
the Poisson brackets (\ref{sum5a})-(\ref{sum5c}) (i.e. the latter coincide 
with the corresponding Dirac brackets), which, in contrast to 
(\ref{sum3a})-(\ref{sum3c}), are completely ultralocal. Furthermore, the 
equations of motion  (\ref{sum9a})-(\ref{sum9b}) simplify to the system 
\begin{eqnarray}
\partial_{t}A_{0} -\partial_{x}A_{1} & = & 0\label{sum6a}\\
\partial_{t}A_{1} -\partial_{x}A_{0} + \frac{1}{t}[A_{1},A_{0}] & = & 0.
\label{sum6b}
\end{eqnarray}
Comparing these equations with the system (\ref{sum1})-(\ref{sum2}) of the 
coset model, one encounters a surprising similarity. However, a term analogous
to the $\frac{1}{t}J_{1}$-term in (\ref{sum2}), which essentially caused the 
coordinate dependence of the spectral parameter of the linear system encoding 
(\ref{sum1})-(\ref{sum2}), is absent. Indeed,  a generalized zero-curvature 
condition with \emph{constant}
spectral parameter can be used  to construct certain non-local charges, which 
are not conserved, but commute with the reduced Hamiltonian \cite{Hus96}.

We now come to the main part of this letter, in which we will  translate  the 
$SL(2,\mathbb{C})$-model into metric variables, thereby revealing its relation
to the coset model. To accomplish this, we follow ref. \cite{NM93} and 
parametrize the vierbein  ${E_{M}}^{A}$ as follows:
\begin{displaymath}
{E_{M}}^{A}=\left(\begin{array}{cc}
N & N^{a}\\
0 & {e_{m}}^{a}
\end{array}\right)\quad\Leftrightarrow \quad {E_{A}}^{M}=
\left(\begin{array}{cc}
N^{-1} &-N^{-1} N^{m}\\
0 & {e_{a}}^{m}
\end{array}\right),
\end{displaymath}
with   $N^{m}\equiv N^{a} {e_{a}}^{m}$, so that the components of the Ashtekar
connection can  be written as
\begin{displaymath} 
A_{ma} = -\frac{1}{4} \varepsilon_{abc}\left( 2\Omega_{dbc}-
\Omega_{bcd}\right) e^{d}_{m} + i e_{mb}\Omega_{0(ab)},
\end{displaymath} where 
\begin{eqnarray*}
{\Omega_{AB}}^{C}& :=& 2{E_{[A}}^{M} {E_{B]}}^{N} \partial_{M} {E_{N}}^{C} = 
-{\Omega_{BA}}^{C}\\
\Omega_{ABC}&:=& {\Omega_{AB}}^{D}\eta_{DC}
\end{eqnarray*}
are the coefficients of anholonomy and the (square-)brackets denote 
(anti-) 
symmetrization containing a factor $1/2$. It is now rather straightforward to 
repeat steps (i) to (iv) at the level of vielbein- (and finally metric-) 
components:

Using adapted coordinates and imposing $\tilde{e}_{\bar{a}}^{x}=
\tilde{e}_{1}^{\bar{m}}=0$ in step (i) leads to 
\begin{eqnarray*}
A\equiv A_{x1} &  = & -\frac{1}{2} \varepsilon_{\bar{b}\bar{c}}
e_{\bar{b}}^{\bar{n}}\partial_{x}e_{\bar{n}\bar{c}} +\frac{i}{N} 
(\partial_{t}e_{x}^{1}-\partial_{x}N^{1})\\
A_{\bar{m}\bar{a}} & = & -\varepsilon_{\bar{a}\bar{b}}e_{1}^{x} 
e_{(\bar{b}}^{\bar{n}}\partial_{x} e_{\bar{n}\bar{c})}\cdot 
e_{\bar{m}}^{\bar{c}} +i\, \frac{1}{N}  e_{\bar{m}\bar{b}}
e_{(\bar{a}}^{\bar{n}}(\partial_{t}-N^{x}\partial_{x}) e_{\bar{n}\bar{b})}\\
A_{x\bar{a}}  & = &- \frac{i}{2N} e_{\bar{n}\bar{a}}\partial_{x}N^{\bar{n}}\\
A_{\bar{m}1} & = &- \frac{i}{2N}e_{1}^{x}e_{\bar{m}}^{\bar{a}} 
e_{\bar{n}\bar{a}}\partial_{x}N^{\bar{n}}.
\end{eqnarray*}
This shows that the conditions $A_{\bar{m}1} =A_{x\bar{a}} = 0$ are equivalent
to $\partial_{x}N^{\bar{n}}=0$. In the case of two-surface orthogonal Killing 
vector fields, one can even choose coordinates such that $N^{\bar{n}}\equiv 0$
so that  the vierbein and therefore the metric become blockdiagonal 
(cf. eq. (\ref{sum0})).

As described in (ii), one now calculates the matrices $B_{\mu}$ and obtains 
for their traceless parts
\begin{eqnarray}
A_{0}& = & -\frac{e_{x}^{1}}{2N}\rho g^{-1}(\partial_{t}-N^{x}\partial_{x})g-
\frac{i}{2}\varepsilon\partial_{x}(\rho g)\label{sum13a}\\
A_{1} & = & -\frac{1}{2}\rho g^{-1}\partial_{x}g-i \frac{e_{x}^{1}}{2N}
\varepsilon(\partial_{t}-N^{x}\partial_{x})(\rho g)\label{sum13b}
\end{eqnarray}
with $\rho g_{\bar{m}\bar{n}}= e_{\bar{m}}^{\bar{a}} e_{\bar{n}\bar{a}}$, 
$\rho=\det(e_{\bar{m}}^{\bar{a}})$ (cf. eq. (\ref{sum0})) and $\varepsilon$ as
in (\ref{sum3d}).

As has already been pointed out in \cite{Hus96}, the final gauge fixings in 
(iv) are nothing but the choice of the (non-stationary) Weyl coordinates 
described at the beginning of this letter. Indeed,  $E\equiv e_{1}^{x}
\det(e_{m}^{a})=\rho$, so that the requirement $t=E$ is equivalent to the 
choice $t=\rho$. On the other hand,  $\Nti=N/(e_{x}^{1}\rho)=N/(e_{x}^{1}t)$, 
so that the consistency condition $\Nti = \frac{i}{tK}$ and the requirement 
$K=i$ imply $N=e_{x}^{1}$, which together with $N^{x}=0$ means that the metric
is of the same form as in (\ref{sum0}) with $t^{-1/2}e^{2k}=(N)^{2}=
(e_{x}^{1})^{2}$. Taking this into account, eqs. (\ref{sum13a})-(\ref{sum13b})
simplify to
\begin{eqnarray}
A_{0} & = & -\frac{1}{2} J_{0}     - \frac{i}{2}\varepsilon \partial_{x}(tg)
\label{sum10a}\\
A_{1} & = & -\frac{1}{2} J_{1}     - \frac{i}{2}\varepsilon \partial_{t}(tg),
\label{sum10b}
\end{eqnarray} 
which is our main result.

It remains to verify whether the system (\ref{sum6a})-(\ref{sum6b}) with 
$A_{\mu}$ as in (\ref{sum10a})-(\ref{sum10b}) is equivalent to the system 
(\ref{sum1})-(\ref{sum2}) and whether the Lagrangian (\ref{sum2a}) really 
reproduces the ultralocal Poisson brackets (\ref{sum5a})-(\ref{sum5c}) for 
combinations such as (\ref{sum10a})-(\ref{sum10b}):

The equivalence of the equations of motion follows from the  identities
\begin{eqnarray*}
\partial_{t}A_{0}-\partial_{x}A_{1} & = & -\frac{1}{2}(\partial_{t} J_{0}-
\partial_{x} J_{1})\\
\partial_{t}A_{1}-\partial_{x}A_{0} +\frac{1}{t} [A_{1},A_{0}] & = & -
\frac{1}{2}\left(\partial_{t}J_{1}-\partial_{x}J_{0}  - \frac{1}{t} 
[J_{1},J_{0}]-\frac{1}{t} J_{1}\right)\nonumber\\
 & & +\frac{i}{2}\varepsilon g\left( -\partial_{t} J_{0}+\partial_{x} 
J_{1}\right),
\end{eqnarray*}
which are valid for any real symmetric and unimodular $(2\times 2)$-matrix 
$g(t,x)$. As for the Poisson brackets, one uses again the method described 
in \cite{KorSam96b} or, alternatively,
the parametrization (\ref{sum2b}) and the corresponding Lagrangian 
(\ref{sum2c}) to infer
\begin{eqnarray*}
\{A_{0}(x)\stackrel{\otimes}{,}A_{0}(x')\} & = &\frac{1}{8C} 
[\Pi,A_{0}\otimes \mathbf{1}]\delta(x-x')\\
\{A_{1}(x)\stackrel{\otimes}{,}A_{0}(x')\} & = &\frac{1}{8C} 
[\Pi,A_{1}\otimes \mathbf{1}]\delta(x-x')\\
\{A_{1}(x)\stackrel{\otimes}{,}A_{1}(x')\} & = & \frac{1}{8C} 
[\Pi,A_{0}\otimes \mathbf{1}]\delta(x-x').
\end{eqnarray*}
Thus, both formulations are  based on the same Poisson structure; it is only 
due to the particular combination of the $(2\times 2)$-matrices in 
(\ref{sum10a})-(\ref{sum10b}) that  all potentially non-ultralocal 
contributions exactly cancel.

This leads to the following conclusions:\\
(i) The $SL(2,\mathbb{C})$-chiral model,  grown out of the Ashtekar 
formulation and equipped with an  ultralocal Poisson structure, and  the 
metric-induced $SL(2,\mathbb{R})/SO(2)$-$\sigma$-model with its 
non-ultralocal Poisson brackets are completely equivalent formulations of the 
two Killing field reduction of general relativity. The  relation between these
two chiral models is displayed by  eqs. (\ref{sum10a})-(\ref{sum10b}), which 
allow it now to translate the results of  one approach into the language of 
the other. As a particularly interesting application one could now  
compare the non-local charges constructed in \cite{KorSam96b} with those 
given in \cite{Hus96}.\\
(ii) The relation (\ref{sum10a})-(\ref{sum10b}) between the currents 
$J_{\mu}$ and $A_{\mu}$ suggests an interesting link to ref. \cite{Byt94}. 
There it has been shown that by a very similar change of variables in the 
$O(3)$-chiral model it is possible to preserve the form of the equations of 
motion while rendering the corresponding  Poisson brackets ultralocal. In our 
case, the transition $J_{\mu}\rightarrow A_{\mu}$ does not \emph{completely} 
preserve  the form of the equations of motion (the $\frac{1}{t} J_{1}$-term 
is eliminated), but also leads to ultralocal Poisson brackets.\\
(iii) While the coset model formulation has  natural generalizations to other 
coset spaces that appear in more complicated models of dimensionally reduced 
(super-)gravity \cite{KorSam96b,KorSam97a}, the $SL(2,\mathbb{C})$-formulation
does not immediately suggest such generalizations, since its existence (like  
the construction in \cite{Byt94}) relies on some peculiarities of 
$(2\times2)$-matrices.\\
\vspace{-3.5mm}\\
The author would like to thank D Korotkin, H Nicolai and   H Samtleben for  
helpful  comments and discussions.\\  


\begin{thebibliography}{99}
\bibitem{Ash86Ash87}A. Ashtekar, \emph{Phys. Rev. Lett.} \textbf{57}, 
2244 (1986); A. Ashtekar \emph{Phys. Rev.} \textbf{D 36}, 1587 (1987)
\bibitem{Hus96}V. Husain, \emph{Phys. Rev.} \textbf{D 53}, 4327 (1996)
\bibitem{HS89}V. Husain and L. Smolin, \emph{Nucl. Phys.} \textbf{B 327}, 
205 (1989)
\bibitem{KorSam96b}D. Korotkin and H. Samtleben, 
\emph{Class. Quantum Grav. }\textbf{14} L151-L156 (1997)
\bibitem{Gow74}R.H. Gowdy, \emph{Ann. of Phys.} \textbf{83}, 203 (1974)
\bibitem{MBZ78}D. Maison, \emph{Phys. Rev. Lett.} \textbf{41}, 521 (1978); 
V.A. Belinskii and V.E. Zakharov, \emph{Sov. Phys. JETP} \textbf{48}, 985 
(1978)
\bibitem{Ger72}R. Geroch, \emph{J. Math. Phys.} \textbf{13}, 394 (1972)
\bibitem{KorSam97a}D. Korotkin and H. Samtleben, gr-qc/9705013 (1997)
\bibitem{Men97}G.A. Mena Marug{\'a}n, \emph{Phys. Rev.} \textbf{D 56} 908 
(1997)
\bibitem{NM93}H. Nicolai and H.-J. Matschull, \emph{Journ. Geom. Phys.} 
\textbf{11}, 15 (1993)
\bibitem{Byt94}A.G. Bytsko,  hep-th/9403101 (1994)
\end{thebibliography}
\end{document}